\documentclass[12pt,preprint]{aastex}

\shorttitle{Which clusters contain central IMBHs}
\shortauthors{Baumgardt et al.}

\begin{document}


\title{Which Globular Clusters contain Intermediate-mass Black Holes?}


\author{Holger Baumgardt\altaffilmark{1},        
        Junichiro Makino\altaffilmark{2},
	Piet Hut\altaffilmark{3}}

\altaffiltext{1}{
        Sternwarte, University of Bonn, Auf dem H\"ugel 71,
        53121 Bonn, Germany}

\altaffiltext{2}{
        Department of Astronomy, University of Tokyo, 7-3-1 Hongo,
        Bunkyo-ku,Tokyo 113-0033, Japan}

\altaffiltext{3}{
        Institute for Advanced Study, Princeton, NJ 08540, USA}


\begin{abstract}

It has been assumed that intermediate-mass black holes (IMBHs) in
globular clusters can only reside in the most centrally concentrated
clusters, with a so-called `core-collapsed' density profile.  While
this would be a natural guess, it is in fact wrong.  We have followed
the evolution of star clusters containing IMBHs with masses between
$125 \le M_{BH} \le 1000$ $M_{\odot}$ through detailed $N$-body
simulations, and we find that a cluster with an IMBH, in projection,
appears to have a relatively large `core' with surface brightness only
slightly rising toward the center.  This makes it highly unlikely that
any of `core-collapsed' clusters will harbor an IMBH.  On the contrary,
the places to look for an IMBH are those clusters that can be fitted
well by medium-concentration King models.

The velocity dispersion of the visible stars in a globular cluster
with an IMBH is nearly constant well inside the apparent core
radius.  For a cluster of mass $M_C$ containing an IMBH of mass
$M_{BH}$, the influence of the IMBH becomes significant only at a
fraction $2.5 M_{BH}/M_C$ of the half-mass radius, deep within the
core, where it will affect only a small number of stars.  In
conclusion, observational detection of an IMBH may be possible, but
will be challenging.

\end{abstract}




\newcommand{\msun}{M_{\odot}}
\def\apgt{\ {\raise-.5ex\hbox{$\buildrel>\over\sim$}}\ }
\def\aplt{\ {\raise-.5ex\hbox{$\buildrel<\over\sim$}}\ }

\section{Introduction}

Over the last few years, four lines of evidence have accumulated
pointing to the possible presence of a $\sim10^3M_\odot$ black hole in
some globular clusters.  The first hint follows from an extrapolation of the
$M_{bh}$--$M_{bulge}$ relation found for super-massive black holes in
galactic nuclei \citep{Magorrianetal1998}, which leads to a prediction
of a typical central black hole mass $\sim10^3M_\odot$ for globular clusters
\citep{KormendyRichstone1995, vanderMarel1999}.
The empirical $M_{bh} - M_C$ relation also comes out naturally      
from rapid mass segregation and the Spitzer instability applied to a standard IMF
in young, dense star clusters \citep{Gurkanetal2004}.

The second hint is related to the discovery of a 
new class of ultra-luminous, compact X-ray sources
(ULXs).  Their high luminosities and strong variability suggest that
they are IMBHs, rather than binaries containing a normal stellar-mass
black hole, and they may occur preferentially in young star clusters
\citep{Zezasetal2002}.

The third hint stems from an analysis of the central velocity
dispersions of specific globular clusters.  \citet{Gerssenetal2002,
Gerssenetal2003} and \citet{Gebhardtetal2002} have published evidence
for black holes in M15 and G1 with masses of order of $10^3$ and $10^4$
$M_\odot$, respectively (since M31's G1 is an order of magnitude more
massive than typical globular clusters in our galaxy, both values
fall on the $M_{bh}$--$M_{bulge}$ relation).

The fourth hint is based on 
detailed $N$-body simulations by \citet{PortegiesZwartetal2004}
of the evolution of a young ($\sim 10$ Myrs) star cluster in M82,
the position of which coincides with an ULX with luminosity $L > 10^{40}$
ergs/sec.  They found that runaway merging of massive stars could have led
to the formation of an IMBH of $\sim10^3M_\odot$.  Since globular
clusters in their youth may have resembled this type of star cluster,
it is altogether likely that at least some globulars harbor IMBHs.

None of these four hints in itself carries enough weight to be convincing.
Given our lack of understanding of the formation process of globulars, there
is no strong reason to expect the $M_{bh}$--$M_{bulge}$ relation to carry
over to globular clusters.  ULXs may just be unusual forms of X-ray binaries
containing a massive but still stellar-mass black hole.  The velocity
dispersion profiles of M15 and G1 can be reproduced by simulations
without central black holes \citep{Baumgardtetal2003a, Baumgardtetal2003b}.
Still, the fact that the four arguments are so different in character does
suggest that we have to take the possible existence of IMBHs very seriously.

The question arises: which globular clusters contain IMBHs?  The intuitive
answer would be: clusters with a steep central luminosity profile,
both because a higher density might suggest an easier formation of a
large black hole, and because such a black hole could be expected to
draw more stars inward.

The main message of this letter is: both of these arguments are wrong.
Since the most plausible formation scenario of an IMBH is run-away
merging in the first $10^7$ years after the formation of a cluster,
dynamical relaxation makes a comparison with current conditions
irrelevant, and thus invalidates the first intuitive argument.  More
importantly, dynamical $N$-body simulations reported in this paper
clearly show that the second intuitive argument is false as well.  We
find that IMBHs, whenever they are formed, quickly puff up the core to
a size far larger than that of the so-called `core collapsed' clusters.
In fact, we show that globular clusters with IMBHs have the appearance
of normal King model clusters, except
in the central regions. We discuss the observational
implications in section 4, after describing our simulation methods in
section 2 and our numerical results in section 3.

\section{Modeling Method}

The reason why clusters with unresolved cores have been the primary
candidate for harboring an IMBH is that there should be a density cusp
with $\rho \propto r^{-7/4}$ around the black hole. The formation of
such a cusp was first predicted by \citet{BahcallWolf1976} and later
confirmed by numerical simulations \citep{CohnKulsrud1978,
MarchantShapiro1980, Baumgardtetal2004a, Pretoetal2004}. The projected density
profile therefore should have a cusp with slope $-3/4$.

A cusp in density, however, does not necessarily imply the existence
of a cusp in luminosity, since there is no guarantee that $M/L$ is
constant. Quite the contrary, numerical simulations of core collapse
have demonstrated that in post-core-collapse clusters $M/L$ shows a
sharp rise toward the center: neutron stars and heavy white dwarfs
dominate the central regions as a result of mass segregation \citep{Baumgardtetal2003a}.  A
similar rise in $M/L$ must exist in the density cusp around an IMBH.

In previous studies, \citet{Baumgardtetal2004a, Baumgardtetal2004b}
have followed the evolution of star clusters with central black holes
with masses of 1 to 10\% of the cluster mass.  They found a distinct
density cusp around the central black hole, but no clear luminosity
cusp, since the central cusp is dominated by remnant stars. The
projected luminosity profile was effectively flat at the center,
and the evolved clusters looked just like normal King clusters.
Unfortunately, the black hole mass used in \citet{Baumgardtetal2004b}
was too large to allow a direct comparison with observations of
globular clusters. In this letter, we report the results of new
simulations, starting with a more realistic central black hole with
a mass of 0.1 to 1\% of the total cluster mass.

The set-up of our runs is similar to that of the runs made by
\citet{Baumgardtetal2004a, Baumgardtetal2004b} and we refer the reader
to these papers for a detailed description.  We simulated the
evolution of star clusters using the collisional $N$-body program
NBODY4 \citep{Aarseth1999} on the GRAPE-6 computers at Tokyo
University. Our simulations include two-body relaxation, stellar
evolution and the tidal disruption of stars by the central black
hole. Initially no binaries were present in our models. Each cluster 
started with a spectrum of stellar masses between
0.1 and 30 $M_\odot$, distributed according to a \citet{Kroupa2001}
mass-function, and massive central black holes initially at
rest at the cluster center. The initial density profile for most
models was given by a King $W_0=7$ model with half-mass radius
$R_H=4.8$ pc.  Stellar evolution was modeled according to the fitting
formulae of \citet{Hurleyetal2000}, assuming a retention fraction of
neutron stars of 15\%. Simulations were run for $T = 12$ Gyrs.

If the $M_{bh}$--$M_{bulge}$ relation found by
\citet{Magorrianetal1998} for galactic nuclei holds for globular
clusters as well, the mass expected for the central black hole in an
average globular cluster of mass $M=1.5 \cdot 10^5 M_\odot$ would be
around 1000 $M_\odot$. Black hole masses of $10^3$ to $3\cdot 10^3 M_\odot$
were also found as the end-result of runaway merging of massive stars
in the dense star cluster MGG-11 by \citet{PortegiesZwartetal2004},
although their values are likely to be upper limits since stellar
mass-loss of the runaway star was not included.  

Since it is not yet
possible to perform a full $N$-body simulation of a massive globular
cluster over a Hubble time, we have to scale down our simulations.
Scaling down can be achieved by simulating either a smaller-$N$
cluster while keeping the mass of the central black hole unchanged, as
done by \citet{Baumgardtetal2004b}, or by scaling down the black hole
mass and the cluster mass simultaneously while keeping the ratio of
both constant. The first method has the advantage that the ratio of
black hole to stellar mass is the same as in a real cluster, allowing
a study of black hole wandering and relaxation processes in the cusp
around the central black hole, while the second method is most
suitable to compare the final velocity and density profile of a star
cluster with observations.  In the present paper we decided to employ
both strategies and made several runs of star clusters containing
$N=65536$ (64K) and $N=131.072$ (128K) cluster stars, and central
black holes with masses of $M_{BH}= 125, 250, 500$ and $1000 M_\odot$
respectively. For a $N=128$K star clusters the final cluster mass is
around $M_C=45.000 M_\odot$, so a cluster with a 250 $M_\odot$ IMBH
would follow the \citet{Magorrianetal1998} relation.

The results of our runs are listed in Table~1 which gives, respectively,
the mass of the black hole, the initial number of cluster stars, the initial
depth of the central potential, the initial half-mass radius, the final
cluster mass, the final half-mass radius, the projected final half-light
radius, the final core radius, the ratio of the last two quantities,
and the logarithm of the final half-mass relaxation time (Spitzer 1987):
\begin{equation}
  T_{rh} = 0.138 \frac{\sqrt{M_C} \; R_H^{3/2}}{<\!m\!> \;\sqrt{G} \;
\ln \Lambda} \;\; ,
\label{trh}
\end{equation}
where $<\!\!m\!\!>$ is the mean mass of all stars in the cluster and
$\ln \Lambda$ is the Coulomb logarithm and is of order 10.  The core
radius was determined as the radius where the surface density has
dropped to half its central value.

\section{Results}

\subsection{Density profile}

We first discuss the density profile of a cluster with an IMBH after
it has evolved for a Hubble time. 
\cite{PortegiesZwartMcMillan2002} and \citet{Rasioetal2004} have shown
that a globular cluster has to
start with a short enough central relaxation time to
form an IMBH by runaway merging of main-sequence
stars. Specific examples have been provided by \cite{PortegiesZwartetal2004},
who found that the star cluster MGG-11 in the starburst galaxy M82,
which has a mass of $3.5 \cdot 10^5 M_\odot$ and an initial half-mass
radius of $R_H=1.2$ pc, could have formed an IMBH while in the
slightly larger cluster MGG-9 with a half-mass radius of $R_H=2.6$ pc,
the time for spiral-in of heavy mass stars was already too long and no
runaway merging occurred.
In order to study the dynamical
evolution of clusters concentrated enough to form IMBHs, we first
simulated two $N=64$K clusters starting with a 3D half-mass radius of
$R_H=2.0$ pc. These clusters have half-mass relaxation times equal to
MGG-11.  Both clusters start with black holes of $125 M_\odot$, in
agreement with the Magorrian relation.  Since the initial density
profile could in principle influence the final density profile and the
dynamical evolution of the cluster, we simulated two clusters starting
from a King $W_0=5$ and a higher concentration $W_0=9$ model
respectively.

\begin{figure}[t]
\plotone{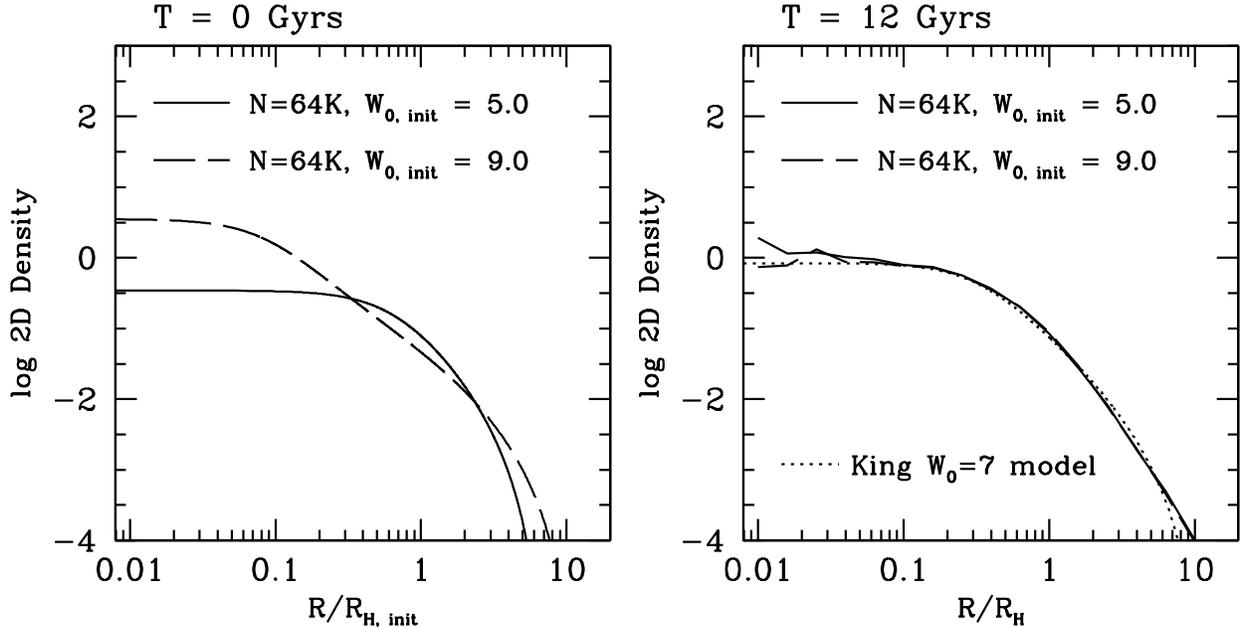}
\caption{2D density profile of bright stars for two $N=64$K clusters starting with half-mass radii
of $r_h=2.0$ pc but different values for the initial central potential $W_0$. The initial relaxation time
was short enough that both clusters expanded by a factor $R_H \sim 5\, R_{H, init}$ and evolved 
toward the same density profile
after a Hubble time. 
\label{fig1}}
\epsscale{1.0}
\end{figure}

Fig.\ \ref{fig1} depicts the projected density profile of bright stars
in both clusters at the start of the simulations and after a Hubble
time.  We defined bright stars to be all giants and all main-sequence
stars with masses larger than 90\% of the mass of turn-off stars. In
order to improve statistics, we overlaid 10 snapshots spaced by 50
Myrs and centered at 12 Gyrs.  Although initially quite different, the
density profiles have become virtually indistinguishable after a
Hubble time.  The reason is that both clusters have expanded strongly:
the final half-mass radii are about 5 to 6 times larger than the
initial ones, and the expansion has erased the initial profile.

As shown in the right panel, this density profile can be fitted rather
well by a King $W_0=7$ model in the range $0.1 \le R/R_H \le 5$.
Outside this range our models show extended halos, which will be
truncated by a background tidal field in most realistic cases.
Inside $R/R_H=0.1$, there is some indication that
the clusters have developed a weak cusp.

\begin{figure}[ht] 
\plotone{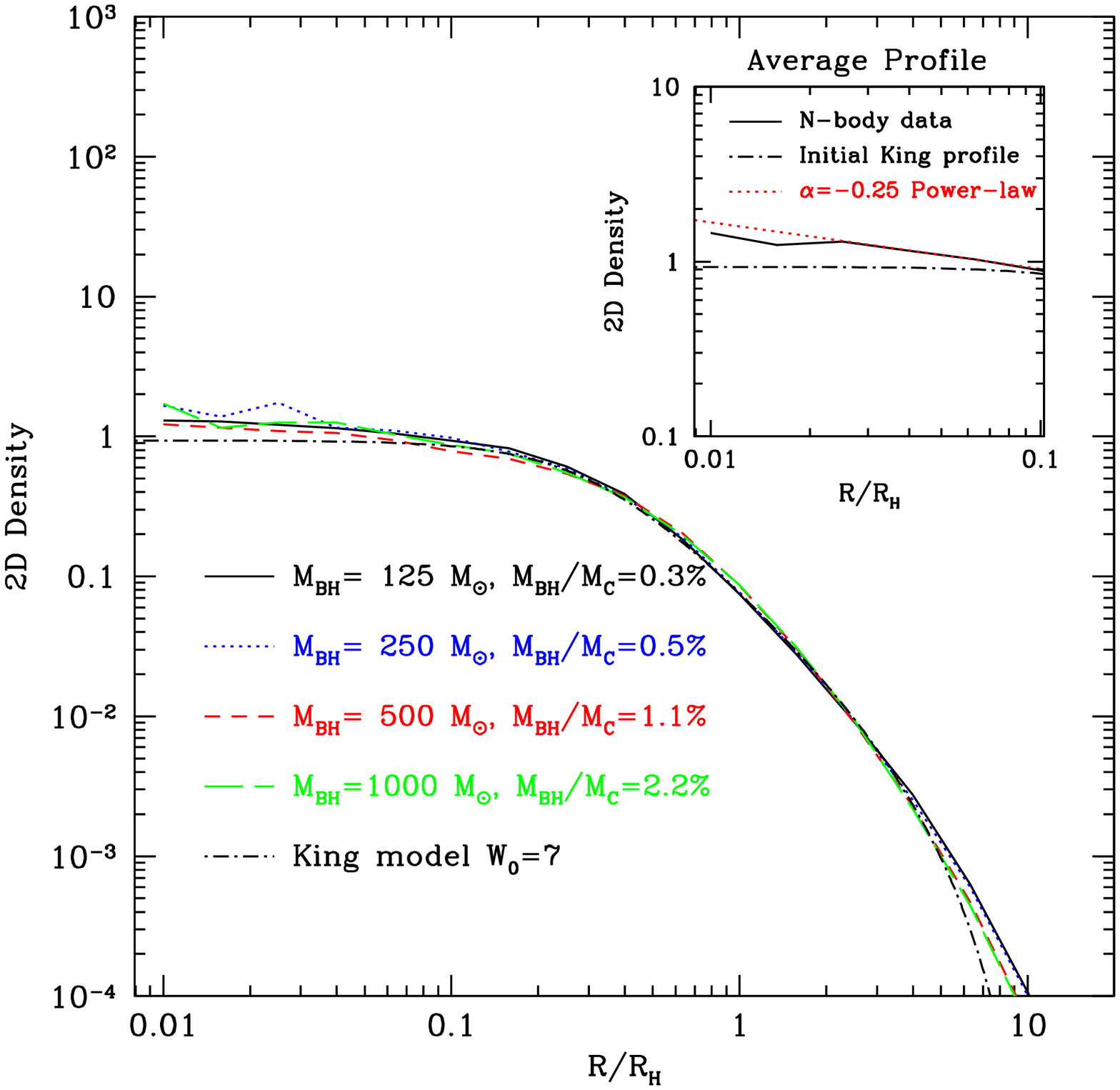}
\caption{Projected density distribution of bright stars after $T=12$ Gyrs
for four
clusters containing black holes between $125 \le M_{BH} \le 1000 M_\odot$
and $N=128$K stars.
All clusters can be fitted by profiles where the density is equal to a King $W_0=7$ model 
outside $R=0.1 R_H$, followed by a density increase in the inner parts. The profiles in
the inner parts are nearly the same for all models. The inset shows the average profile
of all $N$-body runs. Between $0.01 < R/R_H <0.1 $, it can be fitted
by a power-law with slope $\alpha=-0.25$.
This is significantly flatter than the value found for galactic
core-collapsed clusters, $\alpha=-0.8$.
\label{fig2}}
\end{figure}

In order to improve statistics for the inner parts, it was necessary
to add more particles to the simulation. We have simulated a set of
$N=128$K clusters, containing a range of IMBH masses between $125
M_\odot \le M_{BH} \le 1000 M_\odot$.  The starting density profile
was chosen to be a King $W_0=7$ model, close to the equilibrium
profile found above.  These calculations are quite challenging: the
total amount of computing time used for the runs reported in this
letter is more than half of a teraflops-year, or well over $10^{19}$
floating point operations.  It was only through the use of the GRAPE-6
system in the astronomy department of Tokyo University that we were
able to perform these simulations.

Figure \ref{fig2} depicts the projected density distribution of bright
stars after the cluster evolution was simulated for 12 Gyrs.  Between
$0.1 < R/R_H < 3$, the final profiles can be fitted by King $W_0=7$
profiles, and by almost-flat power-law profiles $\rho \sim r^{\alpha}$
inside $R/R_H = 0.1$.  The measured slopes $\alpha$ lie between $-0.1$
and $-0.3$ for the different models, with no clear trend with the mass of
the central black hole. The mean profile of all models
has a slope of $\alpha=-0.25$ (see inset).  A look at
the projected profiles of other stars shows that the heavy mass stars,
i.e. the heaviest white dwarfs and neutron stars, follow significantly
steeper slopes near $-0.5$, reflecting their strong mass segregation.
The overall density profile in the center, however, is quite
close to the density profile of the bright stars since the mass of the
bright stars is close to the average mass of the stars in the core.

According to
\citet{NoyolaGebhardt2004}, slopes of the central surface brightness profiles
of galactic globular 
clusters span a range of values between constant density core models
and models with steep cusps up to $\alpha=-0.8$. The latter value would
correspond to the luminosity
profile expected for a cluster in core-collapse \citep{Baumgardtetal2003a}.
The above results show that core-collapse density profiles are too steep 
for clusters
which contain IMBHs, but that several clusters in the list from 
\citet{NoyolaGebhardt2004} have slopes compatible with the assumptions 
that they contain IMBHs. We will come back to this point in section 4.

\subsection{Velocity dispersion profile}

\begin{figure}[tbp!]
\plotone{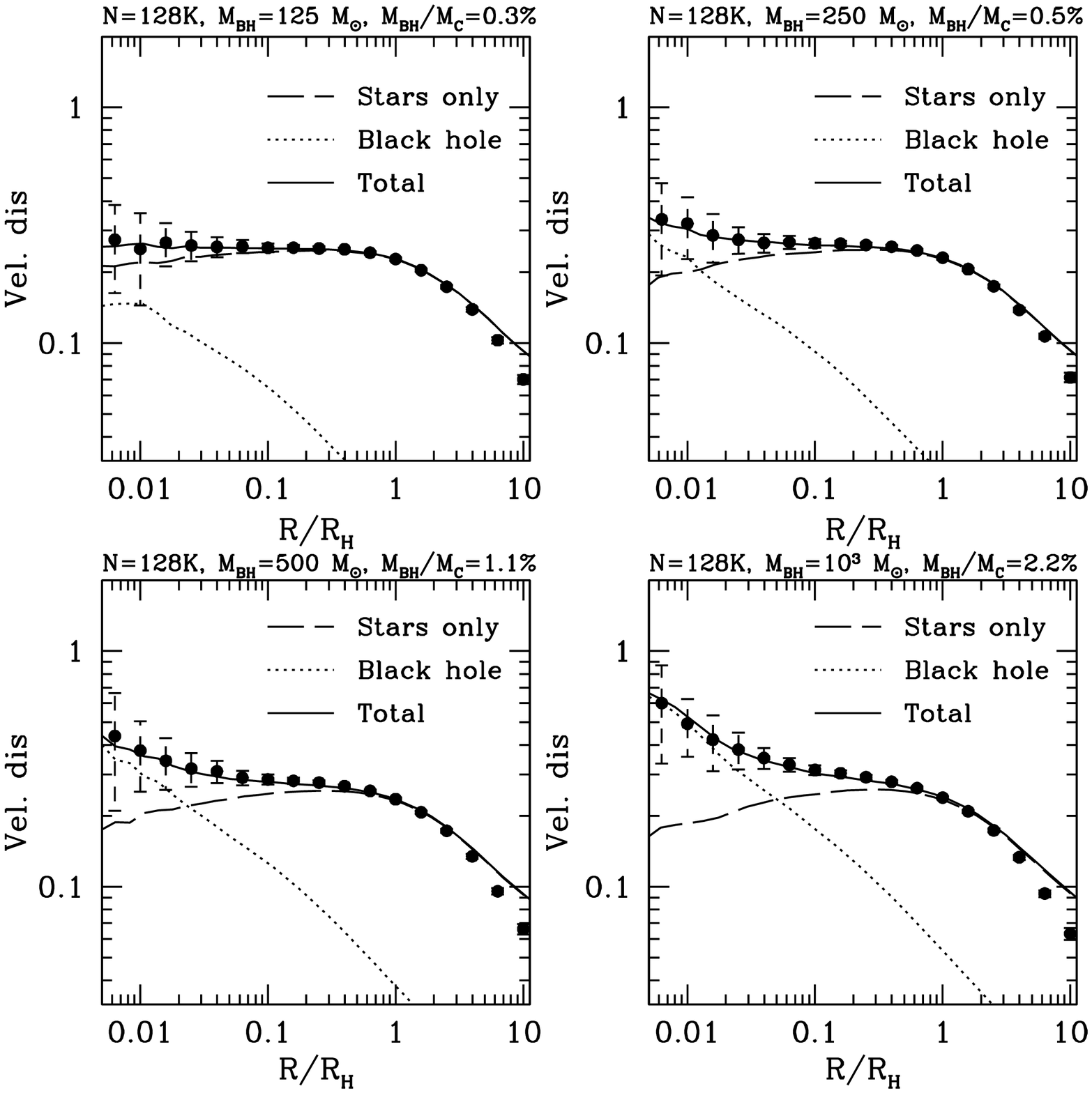}
\caption{Velocity dispersion profiles for four cluster simulations
that started with $N=128$K stars and $M=125, 250, 500$ and $1000
M_\odot$ black holes. Filled circles with error bars are the
velocity dispersion of visible stars in the $N$-body runs. The influence 
of the central black hole grows
with increasing mass.  Also shown are estimates for observational
error bars for a cluster with $5\cdot10^5$ stars in which the
brightest 5\% of all stars can be observed all the way into the
center.
\label{fig3}}
\end{figure}

We next discuss the chances of detecting an IMBH through observations
of the radial velocity or proper motion profiles of a star cluster.
The dots in Fig.~\ref{fig3} show the projected velocity dispersion
profile for four $N=128$K star clusters.  The lines show the predicted
profiles calculated from Jeans equation under the assumption that
the velocity dispersion is isotropic, using as input the potential
from the black hole and the cluster stars.  These predictions form a
very good fit to the $N$-body data, including the region where the
influence of the black hole begins to dominate that of the stars, an
effect that becomes stronger with increasing black hole mass.

Eqs.\ 2 and 3 of \citet{Baumgardtetal2004a} predict a linear relation
between the radius $R$ where the stellar velocity dispersion is
affected more by the IMBH than by the stars alone: $R/R_H = \gamma
M_{BH}/M_{C}$, where $M_C$ is the cluster mass.  The results in Fig.\
3 are compatible with this relation and give $\gamma \approx 2.5$.
The error bars in Fig.\ 3 show the statistical error for a star
cluster containing $5\cdot10^5$ stars and in which the brightest
5\% of all stars can be observed in the center.  For the cluster with
the lowest mass black hole, the black hole dominates only at radii
$R<0.005 R_H$ corresponding to radii of $R<0.5$ arcsec for a typical
globular cluster. There are too few stars inside this radius, so the
velocity error is too large to discern between the black hole and the
no black hole case.  IMBHs with masses $M_{BH}/M_C < 0.3$\% can
therefore not be detected by radial velocity measurements in star
clusters. For the cluster with a $M_{BH}=250 M_\odot$ black hole,
the detection might be possible at the $2\sigma$ level
and higher mass
black holes might be detected even under less favorable conditions.
Nevertheless, even the largest simulated IMBH with 1000 $M_\odot$, which 
has a mass significantly
above the Magorrian et al. relation, creates a central rise that 
is hardly significant if only the brightest cluster stars can be observed. 
Observational detection of an IMBH in a star cluster will therefore be
a challenging task.

\begin{table*}[t]
\caption[]{Results of the $N$-body runs.}
\begin{tabular}{rccccccccc}
\noalign{\smallskip}
\multicolumn{1}{c}{$M_{BH}$} & $N$ & $W_0$ & $R_{Hi}$ & $M_{Cf}$ & $R_{Hf}$ & $R_{HP}$ & $R_C$ & $R_C/R_{HP}$ 
  & lg $T_{RH}$ \\ 
\multicolumn{1}{c}{[$M_\odot$]} & & &   [pc] & [$M_\odot$] & [pc] & [pc] & [pc] &  & [yrs] \\ 
\noalign{\smallskip}
 125 &  65536 & 5 & 2.03 & 21451.3 & 10.12 & 4.14 & 0.55 & 0.13 & 9.54 \\ 
 125 &  65536 & 9 & 2.03 & 21749.6 & 12.34 & 6.20 & 0.65 & 0.11 & 9.67 \\
 125 & 131072 & 7 & 4.91 & 45534.8 & 12.31 & 5.98 & 0.81 & 0.14 & 9.82 \\
 250 & 131072 & 7 & 4.91 & 45311.2 & 12.60 & 6.46 & 0.71 & 0.11 & 9.84 \\
 500 & 131072 & 7 & 4.91 & 44771.1 & 13.70 & 7.76 & 0.64 & 0.08 & 9.89 \\
1000 & 131072 & 7 & 4.91 & 45300.4 & 14.07 & 7.96 & 0.58 & 0.08 & 9.91\\
\end{tabular}
\end{table*}

\section{Galactic globular cluster candidates}

In this section we will compare the projected density profile of bright stars in our simulations
with the observed central surface brightness profiles of galactic globular clusters.
\citet{NoyolaGebhardt2004} have determined surface brightness profiles for 37 globular clusters
from previously published HST WFPC2 images. They found that the slopes of central surface 
brightness profiles
follow a range of values, from 0 (i.e. flat cores) to $-0.8$.
As was shown in section 3.1, the most
promising candidate clusters for IMBHs have central surface brightness slopes of $-0.25$ and outer
profiles that can be fitted by King models with $W_0=7$, corresponding to a concentration 
parameter $c=1.5$.
Slightly different values for $W_0$ and $c$ might be possible if the tidal field
plays an important role and removes stars in the halo.

Table~2 lists all clusters whose profile is compatible with a central slope between $-0.2$ and
$-0.3$ and incompatible with a flat core from the list
of \citet{NoyolaGebhardt2004}. We have also listed the central concentration $c$ of the clusters,
the projected half-light radii as given by \citet{Trageretal1995} and the core radii as determined
by \citet{NoyolaGebhardt2004}. Core and half-light
radii were transformed into physical units with the cluster distances from \citet{Harris1996}.
The final column gives the half-mass
relaxation times, calculated from the cluster masses 
and the half-light radii, assuming that the (3D-)half-mass radius is twice as large
as the (2D-)half-light radius. This is approximately the case in our runs. 

It can be seen that a total of 9 clusters out of 37 have central surface-brightness slopes in
agreement with our simulations. Among these, NGC 6397 is an unlikely candidate since the central
slope is rather steep and its concentration $c$ is more compatible with a core-collapsed cluster.
The same could be true for NGC 5824 and NGC 6541.
The half-mass relaxation time of NGC 6715 is rather long compared to what our clusters reach
after a Hubble time. For the remaining 5 clusters, the central slopes, the ratio of the core to the
half-light radius and the relaxation times are in good
agreement of what we would predict for a cluster with an IMBH.  It
would therefore be extremely interesting to obtain accurate radial
velocity dispersions for these clusters in order to either detect IMBHs or place
upper limits on their possible mass.

\begin{table*}[t]
\caption[]{GC candidates which could contain IMBHs from the list of Noyola \& Gebhardt (2004)}

\begin{tabular}{ccccccccc}
\noalign{\smallskip}
Name & Central & \multicolumn{1}{c}{$\log{M_C}$} & $R_{HP}$ & $R_C$ & $R_C/R_{HP}$ & c & $\log{T_{rh}}$ \\ 
 & Slope & [$M_\odot$] & [pc] & [pc] & & & [yrs] \\ 
   NGC 5286  & $-0.20 \pm 0.02$ & 5.67 & 2.44 & 0.18 & 0.08 & 1.46 &  9.72 \\ 
   NGC 5694  & $-0.21 \pm 0.10$ & 5.35 & 3.28 & 0.34 & 0.10 & 1.84 &  9.76 \\ 
   NGC 5824$^1$  & $-0.38 \pm 0.08$ & 5.15 & 3.35 & 0.20 & 0.06 & 2.45 &  9.67 \\
   NGC 6093  & $-0.13 \pm 0.04$ & 5.51 & 1.89 & 0.24 & 0.13 & 1.95 &  9.48 \\
   NGC 6266  & $-0.15 \pm 0.04$ & 5.90 & 1.92 & 0.20 & 0.08 & 1.70 &  9.68 \\ 
   NGC 6388  & $-0.14 \pm 0.03$ & 5.99 & 1.53 & 0.20 & 0.10 & 1.70 &  9.58 \\ 
   NGC 6397$^1$  & $-0.29 \pm 0.03$ & 4.87 & 1.94 & 0.03 & 0.02 & 2.50 &  9.17 \\ 
   NGC 6541$^1$  & $-0.36 \pm 0.07$ & 5.56 & 2.42 & 0.13 & 0.05 & 2.00 &  9.67 \\
   NGC 6715$^1$  & $-0.16 \pm 0.07$ & 6.23 & 3.58 & 1.30 & 0.34 & 1.84 & 10.26 \\[+0.3cm] 
\end{tabular}
\begin{flushleft}
Notes: 1: Unlikely to contain IMBHs, see text
\end{flushleft}
\end{table*}

\section{Conclusions}

We have followed the evolution of star clusters containing central
IMBHs with masses $125 \le M_{BH} \le 1000 M_\odot$.  All clusters
show a final density profile corresponding to a King $W_0=7$ model
outside the cluster core.  Inside the core, the projected distribution
of bright stars is almost flat, with only a weak rise toward the
center in the form of a power-law of slope $\alpha \sim -0.25$.  We
conclude that the luminosity profiles of several galactic globular
clusters are in good agreement with the assumption that they contain
IMBHs.

A definite detection of an IMBH in a globular cluster can only be made
through observations of the velocity dispersion profile of stars deep
within the core, since the radius where the influence of the black
hole dominates over the cluster stars is given by $R/R_H=2.5
M_{BH}/M_C$.  This radius is an order of magnitude smaller than the
core radius, if the IMBH mass follows the Magorrian relation.  About
25 stars would have to be observed inside this radius to detect the
black hole at a $2\sigma$-level.

\section*{Acknowledgments}

We thank Karl Gebhardt for sending us his draft prior to publication.
We also thank the referee Fred Rasio for comments which improved the
presentation of the paper.


\begin{thebibliography}{}
\bibitem[Aarseth(1999)]{Aarseth1999}
Aarseth, S. J. 1999, \pasp, 111, 1333
\bibitem[Bahcall \& Wolf(1976)]{BahcallWolf1976}
Bahcall, J. N. and Wolf, R. A. 1976, \apj, 209, 214
\bibitem[Baumgardt et al.(2003a)]{Baumgardtetal2003a}
Baumgardt, H., Hut, P., Makino, J., McMillan, S. and Portegies Zwart S.
2003a, \apj, 582, L21
\bibitem[Baumgardt et al.(2003b)]{Baumgardtetal2003b}
Baumgardt, H., Makino, J., Hut, P., McMillan, S. and Portegies Zwart S.
2003b, \apj, 589, L25
\bibitem[Baumgardt, Makino \& Ebisuzaki(2004a)]{Baumgardtetal2004a}
Baumgardt, H., Makino, J., and Ebisuzaki, T. 2004a, \apj, 613, 1133
\bibitem[Baumgardt, Makino \& Ebisuzaki(2004b)]{Baumgardtetal2004b}
Baumgardt, H., Makino, J., and Ebisuzaki, T. 2004b, \apj, 613, 1143 
\bibitem[Cohn \& Kulsrud(1978)]{CohnKulsrud1978}
Cohn, H., and Kulsrud, R. M. 1978, \apj, 226, 1087
\bibitem[Gebhardt et al.(2000)]{Gebhardtetal2000}
Gebhardt, K., et al. 2000, \apj, 539, L13
\bibitem[Gebhardt et al.(2002)]{Gebhardtetal2002}
Gebhardt, K., Rich, R. M., and Ho, L. C. 2002, \apj, 578, L41
\bibitem[Gerssen et al.(2002)]{Gerssenetal2002}
Gerssen, J., van der Marel, R. P., Gebhardt, K., Guhathakurta, P.,
Peterson R. C., and Pryor, C., 2002, \aj, 124, 3270
\bibitem[Gerssen et al.(2003)]{Gerssenetal2003}
Gerssen, J., van der Marel, R. P., Gebhardt, K., Guhathakurta, P.,
Peterson R. C., and Pryor, C., 2003, \aj, 125, 376
\bibitem[G\"urkan et al.(2004)]{Gurkanetal2004}
G\"urkan, M. A., Freitag, M., Rasio, F., 2004, \apj, 604, 632
\bibitem[Harris(1996)]{Harris1996}
Harris, W. E. 1996, \aj, 112, 1487 ({\tt
http://physun.physics.mcmaster.ca/Globular.html})
\bibitem[Hurley et al.(2000)]{Hurleyetal2000}
Hurley, J. R., Pols, O. R., and Tout, C. A. 2000, \mnras, 315, 543
\bibitem[Magorrian et al.(1998)]{Magorrianetal1998}
Magorrian, J., et al. 1998, \aj, 115, 2285
\bibitem[Marchant \& Shapiro(1980)]{MarchantShapiro1980}
Marchant, A. B., and Shapiro, S. L. 1980, \apj, 239, 685
\bibitem[Matsumoto et al.(2001)]{Matsumotoetal2001} 
Matsumoto, H., et al. 2001, \apj, 547, L25
\bibitem[Miller \& Hamilton(2002)]{MillerHamilton2002}
Miller, M. C., and Hamilton D. P. 2002, \mnras, 330, 232
\bibitem[Mouri \& Taniguchi(2002)]{MouriTaniguchi2002} 
Mouri, H., and Taniguchi, Y. 2002, \apj, 566, L17
\bibitem[Noyola \& Gebhardt(2004)]{NoyolaGebhardt2004}
Noyola, E., Gebhardt, K., 2004, preprint
\bibitem[Kaaret et al.(2001)]{Kaaretetal2001} 
Kaaret, P., et al. 2001, \mnras, 321, L29
\bibitem[Kalogera, King \& Rasio(2004)]{Kalogeraetal2004}
Kalogera, V., King, A. R., Rasio, F. A., 2004, \apj, 601, L171
\bibitem[Kormendy \& Richstone(1995)]{KormendyRichstone1995}
Kormendy, J., Richstone, D., 1995, \araa, 33, 581
\bibitem[Kroupa(2001)]{Kroupa2001}
Kroupa, P. 2001, \mnras, 322, 231
\bibitem[Portegies Zwart \& McMillan(2002)]{PortegiesZwartMcMillan2002}
Portegies Zwart, S. F., and McMillan, S. L. W. 2002, \apj, 576, 899
\bibitem[Portegies Zwart et al.(2004)]{PortegiesZwartetal2004}
Portegies Zwart, S. F., Baumgardt, H., Hut, P., Makino, J., and McMillan, S. L. W.
2004, \nat, 428, 724 
\bibitem[Preto, Merritt \& Spurzem (2004)]{Pretoetal2004}
Preto, M., Merritt, D., Spurzem, R. 2004, in preparation, astro-ph/0406324
\bibitem[Rasio et al.(2004)]{Rasioetal2004} 
Rasio, F. A., Freitag, M., and G\"urkan, M. A. 2004, in {\it Coevolution of
Black Holes and Galaxies}, Carnegie Observatories Astrophysics Series I,
ed. L. C. Ho, p.\ 138
\bibitem[Trager et al.(1995)]{Trageretal1995}
Trager, S. C., King, I., Djorgovski, S. 1995, \aj, 109, 218
\bibitem[van der Marel(1999)]{vanderMarel1999}
van der Marel, R. P., 1999, in {\it Black Holes in Binaries and
Galactic Nuclei}, Proceedings of the ESO Workshop held at Garching, Germany,
eds. E. P. J. van den Heuvel, P. A. Woudt, p.\ 246
\bibitem[Zezas et al.(2002)]{Zezasetal2002}
Zezas, A., Fabbiano, G., Rots, A. H., Murray, S. S. 2004, \apj, 577, 710 
\end{thebibliography}
\end{document}